\documentclass[11pt,preprint]{aastex}
\def\lta{\lower2pt\hbox{$\buildrel {\scriptstyle <}
   \over {\scriptstyle\sim}$}}
\def\gta{\lower2pt\hbox{$\buildrel {\scriptstyle >}
   \over {\scriptstyle\sim}$}}
\def\mfb{\dot{m}_{\rm fb}}
\def\mbh{\dot{m}_{\rm BH}}
\def\macc{\dot{m}_{\rm acc}}
\def\rfb{r_{\rm fb}}
\def\rd{r_d}

\shorttitle{GRB Central Engine}

\begin{document}

\title{Mass Fall-back and Accretion in the Central Engine of Gamma-Ray Bursts}

\author{Pawan Kumar$^1$, Ramesh Narayan$^2$ \& Jarrett L. Johnson$^1$ \\
$^1$Astronomy Department, University of Texas, Austin, TX 78712\\
$^2$Harvard-Smithsonian Center for Astrophysics, 60 Garden Street, Cambridge, MA 02138
}

\begin{abstract}
We calculate the rate of in-fall of stellar matter on an accretion
disk during the collapse of a rapidly rotating massive star, and
estimate the luminosity of the relativistic jet that results from
accretion on to the central black hole.  We find that the jet
luminosity remains high for about $10^2$ s, at a level comparable to
the typical luminosity observed in gamma-ray bursts (GRBs).  The
luminosity then decreases rapidly with time for about $\sim10^3$ s,
roughly as $\sim t^{-3}$; the duration depends on the size and rotation
speed of the stellar core.  The rapid decrease of the jet power
explains the steeply declining x-ray flux observed at the end of most
long duration GRBs.

Observations with the Swift satellite show that, following the steep
decline, many GRBs exhibit a plateau in the x-ray lightcurve (XLC)
that lasts for about 10$^4$ s.  We suggest that this puzzling feature
is due to continued accretion in the central engine.  A plateau in the
jet luminosity can arise when the viscosity parameter $\alpha$ is
small, $\sim10^{-2}$ or less. A plateau is also produced by continued
fall-back of matter -- either from an extended stellar envelope or
from material that failed to escape with the supernova ejecta. In a
few GRBs, the XLC is observed to drop suddenly at the end of the
plateau phase, while in others the XLC declines more slowly as $\sim
t^{-1}-t^{-2}$.  These features arise naturally in the accretion model
depending on the radius and mean specific angular momentum of the
stellar envelope.

The total energy in the disk-wind accompanying accretion is found to
be about 10$^{52}$ erg.  This is comparable to the energy observed in
supernovae associated with GRBs, suggesting that the wind might be the
primary agent responsible for the explosion.

The accretion model thus provides a coherent explanation for the diverse
and puzzling features observed in the early XLC of GRBs.  It might be
possible to use this model to invert gamma-ray and x-ray observations
of GRBs and thereby infer basic properties of the core and envelope of
the GRB progenitor star.

\end{abstract}

\section{Introduction}

Observations of gamma-ray bursts (GRBs) carried out by the NASA Swift
satellite in the last two years have shown that the $\gamma$-ray
prompt emission turns off abruptly after about a minute.  The abrupt
shutoff is evidenced by the rapidly declining x-ray flux (t$^{-3}$ or
faster) from about 80s to 300s, which joins smoothly the prompt GRB
lightcurve when extrapolated back in time and spectral band
(Tagliaferri et al. 2005; Nousek et al. 2006; O'Brien et
al. 2006). The abrupt decline suggests that the activity at the center
of the explosion declines very rapidly with time after about a minute
of more or less steady activity.

At the same time, Swift observations have provided overwhelming
evidence that the GRB central engine continues operating for hours and
perhaps even days. There are two independent indications for this
phenomenon.  First, we often see a phase in the early x-ray lightcurve
of long-GRBs, from $\sim 10^3$s to 10$^4$s, during which the flux
declines slowly with time.  We will refer to this as a ``plateau'' in
the lightcurve.  There are many proposals to explain the plateau. 
The models are, however, highly constrained by a lack of correlation
between x-ray and optical features for many GRBs.  Therefore, 
a successful proposal must
invoke continuing activity of the central engine to produce the x-ray plateau (e.g.,
Zhang 2006; Panaitescu 2007, and references therein), whereas the 
simultaneous optical emission may be produced in the afterglow.

Second, in roughly a third of the observed GRBs, the x-ray flux is
seen to increase suddenly and then to drop precipitously in what are
referred to as ``flares.''  In some cases the flux during these flares
increases by a factor of $\sim 10^2$ on a time scale $\delta t\ll t$
(Burrows et al. 2005; Chincarini et al. 2007), which cannot be
explained in terms of a density inhomogeneity in the external medium
(Nakar \& Granot, 2006).  Variable activity in the central engine is a
more natural explanation.  A sudden drop in the flux --- e.g., in the
case of GRB 070110 the x-ray lightcurve was nearly flat for 20 ks and
then fell off as $t^{-8}$ (Troja et al. 2007) --- is also not possible
to understand other than as the result of highly variable central
engine activity.

The above conflicting requirements, viz., (i) a sudden drop in
activity at the end of the main GRB ($t\ \lta 10^2$ s), (ii) continued
steady activity during an extended plateau ($t\sim10^4$ s), and (iii)
occasional dramatic flares, are challenging for models of the central
engine.  In this paper we consider the currently most popular model of
GRBs, which postulates ultra-rapid accretion of gas on to a
newly-formed black hole or neutron star (Narayan, Paczy\'nski \& Piran
1992; Popham, Woosley \& Fryer 1999; Narayan, Piran \& Kumar 2001).
The accretion disk may be the result of (i) gas fall-back after a
hypernova, as in the collapsar model (Woosley 1993; Paczy\'nski 1998;
MacFadyen \& Woosley 1999), which is considered relevant for long duration
GRBs, or (ii) during the merger of a double neutron star or neutron
star-black hole binary (Eichler et al. 1989; Narayan et al. 1992),
which is currently considered the most likely explanation for short duration
GRBs. We attempt to reconcile the accretion model for collapsars with 
Swift observations of long duration GRBs.  

In \S~2, we make use of a reasonably realistic model of the progenitor
of a collapsar and estimate the rate at which gas is added to the
accretion torus at the center.  The calculation is based on a crude
free-fall model for the collapsing star and the results are to be
taken in the spirit of an order of magnitude estimate.
Then, in \S~3, we use this model to study the variation of accretion
power with time.  We show that the model naturally reproduces both the
sudden shutoff of the prompt GRB emission and the extended
plateau in the lightcurve.  We speculate on possible scenarios for
producing flares and suggest an explanation for the hypernova
explosion associated with some GRBs.  We conclude in \S~4 with a
discussion and summary.

\section{Mass Fall-back Rate}

\subsection{Particle Trajectories}

Consider an axisymmetric rotating star at the onset of core
collapse. For simplicity, let us ignore pressure forces and take the
trajectory of each particle to correspond to free-fall. At the
beginning of the collapse the velocity field is in the $\hat\phi$
direction, and therefore particles start out at apo-center. A particle
that is initially located at $(r,\theta,\phi)$ with angular velocity
$\Omega(r,\theta)$ will follow a trajectory with semimajor-axis $a$
and eccentricity $e$ given by
\begin{equation}
a = {r\Omega^2\sin^2\theta\over\Omega_k^2 (1-e^2)} = {r\over 1+e}, \quad\quad
e = 1 - {\Omega^2\over \Omega_k^2}\sin^2\theta,
\label{ae}
\end{equation}
and its coordinates will vary with time $t$ as
\begin{eqnarray}
x(t) &=& a(e+\cos\eta)\sin\theta\cos\phi + a (1-e^2)^{1/2}\sin\eta\sin\phi, 
    \\\nonumber\\[-.5cm]
y(t) &=& a(e+\cos\eta)\sin\theta\sin\phi - a (1-e^2)^{1/2}\sin\eta\cos\phi, 
   \\\nonumber\\[-.5cm]
z(t) &=& a(e+\cos\eta)\cos\theta.
\end{eqnarray}
Here $\eta$ is related to $t$ via
\begin{equation}
t = \Omega_k^{-1} (\eta + e\sin\eta)(1+e)^{-3/2},\quad\quad\quad\quad \Omega_k =
  \left({G M_r\over r^3}\right)^{1/2},
\end{equation}
where $M_r$ is the mass enclosed within radius $r$, and $\eta=0$ at
$t=0$,

The particle trajectory intersects the equatorial plane when
$\cos\eta=-e$, or
\begin{equation}
t_{\rm eq} = \Omega_k^{-1}\left[\cos^{-1}(-e) + e(1-e^2)^{1/2}
   \right](1+e)^{-3/2} + t_s(r),
\label{teq}
\end{equation}
and the distance of the particle from the center at this time is
\begin{equation}
r_{\rm eq}(r,\theta)=r(1-e).
\label{req}
\end{equation}
The term $t_s(r)$ in equation (\ref{teq}) is the sound travel time
from the center to radius $r$, which is roughly the time it takes
(from the start of collapse at the center) for gas at $r$ to realize
the loss of pressure support and to begin its fall toward the center.
Apart from ignoring pressure support during the collapse, the above
analysis ignores a number of other effects, e.g., a wind from the
accretion disk which might inhibit fall-back, a shock generated by the
bounce provided by a neutron star if one is created in the initial
collapse, etc. Therefore, the results we present in this paper 
might have an error of a factor of a few. The purpose of this work is to
identify a possible cause for the rapid decrease in central engine
activity and the subsequent plateau and flare emission, and to help
understand the dependence of accretion rate on stellar structure and
rotation profile. It might perhaps also be useful for understanding
certain aspects of 3-D numerical simulations of collapsars.

\subsection{Formation of an Accretion Disk}
As particles intersect the equatorial plane during their free-fall,
they will become part of a thick accretion disk that is centrifugally
supported around the nascent black hole, provided they possess
sufficient angular momentum.  The formation of an accretion disk
is likely required for launching a jet from an accreting black hole
and producing a long duration GRB via the collapsar model
(e.g. MacFadyen \& Woosley 1999); however, see Proga (2005) for an
alternate point of view. Clearly, it is important to
determine the conditions under which an accretion disk will form.

Given axisymmetry of the progenitor star about the rotation axis and
mirror symmetry across the equator, when a particle hits the
equatorial plane from above it will collide with another particle with
the opposite sign of $v_z$ that started at the mirror location below
the plane (Fig. 1).  The velocities of the two particles when they
collide at the equatorial ($x$-$y$) plane are given by
\begin{equation}
v_x = -{r\Omega_k\left[\sin\theta\cos\phi+e\sin\phi\right]\over(1-e)^{1/2}},
   \quad v_y = -{r\Omega_k\left[\sin\theta\sin\phi - e\cos\phi \right] 
   \over(1-e)^{1/2}}, \quad v_z=\mp{r\Omega_k\cos\theta \over(1-e)^{1/2}}.
\end{equation}
Collisions of gas blobs are highly inelastic, and so the two particles
will merge and result in zero $z$-velocity. The dissipation of the energy
associated with this component of the velocity will heat up the gas
to nearly the virial temperature, and therefore the geometry of the infalling
gas when it reaches the central region starts out as a thick disk (the disk
thickness can become smaller at smaller radius if neutrino cooling is 
efficient). The surviving fluid velocity in the equatorial plane can 
be decomposed into azimuthal (circulation) and radial (in-fall) components:
\begin{equation}
v_\phi = -{r\Omega_k\sin\theta\over (1-e)^{1/2}}, \quad\quad
  v_r = -{r\Omega_k e\over (1-e)^{1/2}}.
\end{equation}

\begin{figure}[h!]
\begin{center}
\includegraphics[width=11.0cm,height=9.0cm]{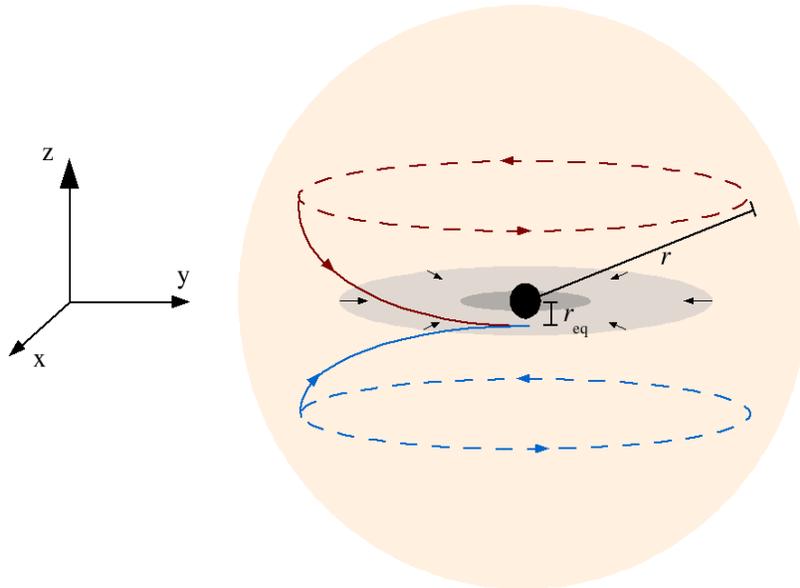}
\caption{\small The formation of a thick accretion disk around the black hole 
  at the center of a collapsing star.  A parcel of gas which has a
  circular orbit about the axis of rotation of the star at a radius
  $r$ before core collapse (red dashed line), goes into free-fall
  after core collapse (solid red line), colliding with its counterpart
  from the opposite side of the equatorial plane (blue dashed and
  solid lines) at a distance $r_{\rm eq}$ and forming a disk in the
  equatorial (x-y) plane (light gray).  After the parcels of gas
  collide, they fall further toward the black hole (black) until their
  angular momentum allows them to become centrifugally supported in an
  accretion disk (dark gray).  The set of axes is shown for
  orientation; in our calculations, the origin of this coordinate
  system is at the center of the star.}
\end{center}
\end{figure}

The velocity field immediately following the in-fall of stellar matter
on the equatorial plane is convergent toward the center of the
star --- the sign of $v_r$ is negative, independent of the initial
particle position, and $|v_r|\ \gta\ v_\phi$. This leads to a rapid
shrinking of the initial torus formed in the stellar collapse, as
shown in Fig. 1. The rapid inward flow is terminated at a radius where
the specific angular momentum of the in-falling gas is equal to the
angular momentum of a locally circular orbit. We note that the mean
specific angular momentum of the in-falling gas when it first hits the
equatorial plane, $\ell = \left< v_\phi r(1-e) \right>$, is smaller
than the angular momentum needed for a circular orbit at that radius
($r_{\rm eq} = \left< r(1-e)\right>$) by a factor of about $3\pi/8$, 
where $\left< \right>$ denotes angular
averaging. Thus, the initial torus radius will shrink by the square of
this factor, i.e., by a factor of $\sim$ 1.4, before the gas becomes
centrifugally supported. Thus, the radius at which the
material goes into quasi-circular orbit, which we refer to as the
fall-back radius $r_{\rm fb}$, is approximately given by
\begin{equation}
r_{\rm fb} \approx {r_{\rm eq}\over 1.4} =
{\left< r(1-e)\right>\over 1.4}.
\label{rfb}
\end{equation}

The descend from $r_{\rm eq}$ to $r_{\rm fb}$, a 30\% drop in orbital
radius, takes place on
a local free-fall timescale, which is much shorter than $t_{\rm eq}$,
and the ratio of the initial and final radius within the equatorial
plane is almost independent of the initial particle position.
Therefore, we take the rate at which gas settles into a centrifugally
supported accretion disk as being the same as the rate at which
stellar matter lands on the equatorial plane.  The latter is obtained
from equations (\ref{teq}) and (\ref{req}), which map the initial
particle position to the position on the disk at which the particle
intersects the equatorial plane (note that
$\phi\longrightarrow\phi-\pi/2$). The Jacobian of the
transformation $(r,\theta,\phi)\longrightarrow (r_{\rm eq},\theta\rightarrow
\pi/2, \phi-\pi/2)$ gives the mass fall-back rate per unit area on the disk.

\begin{figure}[h!]
\begin{center}
\includegraphics[width=5.95in,height=3.9in]{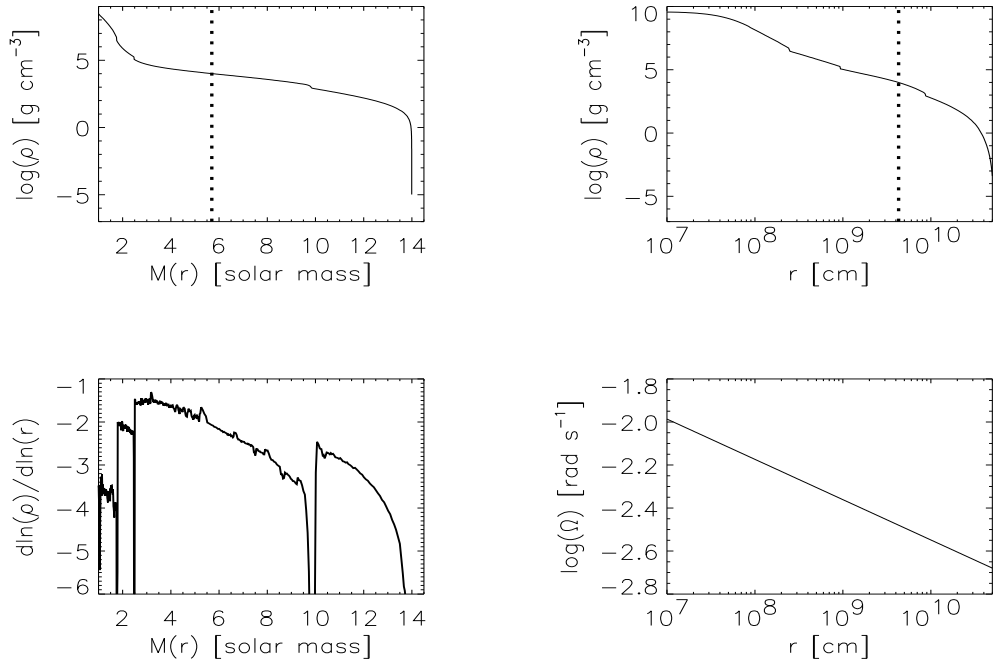}
\caption{\small The top left panel shows the density $\rho$ of a 14
  M$_{\odot}$ GRB progenitor star (model 16T1 of Woosley \& Heger,
  2006), as a function of enclosed mass $M(r)$, and the top right
  panel shows $\rho$ as a function of $r$. The lower left panel shows
  $\tau\equiv d\ln\rho/ d\ln r$, as a function of $M(r)$ (sharp
  glitches in $\tau$ are associated with composition changes), and the
  lower-right panel shows the angular velocity $\Omega$ ($\Omega$ is 
  NOT from the evolutionary calculation of Woosley \& Heger, 2006). The 
  vertical dotted line shows the part of the stellar core that falls 
  directly to form a black hole; the region outside of this has sufficiently
  large specific angular momentum to form a disk. }
\end{center}
\end{figure}

\subsection{Numerical Results}

The rate at which material falls back on the
accretion disk depends on both the density profile of the star and on
its angular velocity profile. In this paper,
we use a pre-GRB-collapse stellar model developed by Woosley
and Heger (2006) for their collapsar simulations (model 16T1); the
mass and radius of this progenitor star are 14 M$_{\odot}$ and
5.18x10$^{10}$cm (the core can be modeled as a polytrope of index
4.5). Figure 2 shows the density profile for this star.  Our choice of
the rotation profile for the model (Fig. 2, lower right panel) is guided by the
evolution calculation of Rockefeller et al. (2006).

In Fig. 3 we show for the above stellar model the rate at which mass
is added to the disk, and the mean radius at which the gas goes into a
circular orbit. In these calculations we assumed that mass fall-back
occurs only within a wedge extending $\pm60^o$ from the equatorial
plane.  This allows crudely for the effect of a wind from the central
disk which might prevent fall-back along the polar regions.  We see
that gas inflow on the disk starts out at a high rate and occurs
initially at a small radius. With increasing time the rate decreases
rapidly ($\sim t^{-3}$), and much of the gas falls at a larger and
larger distance from the center. Material 
that lands on the equatorial plane at times earlier than about 20 seconds
after core collapse falls directly into the black hole, whereas
gas falling later than 20 seconds becomes centrifugally supported
outside the Schwarzschild radius of the growing black hole.
Therefore, we expect our model progenitor star to form an accretion
disk at roughly this time after the initiation of core collapse.  This
is presumably also the time when a jet first forms.

\begin{figure}[h!]
\begin{center}
\includegraphics[height=4.3in,width=5.85in]{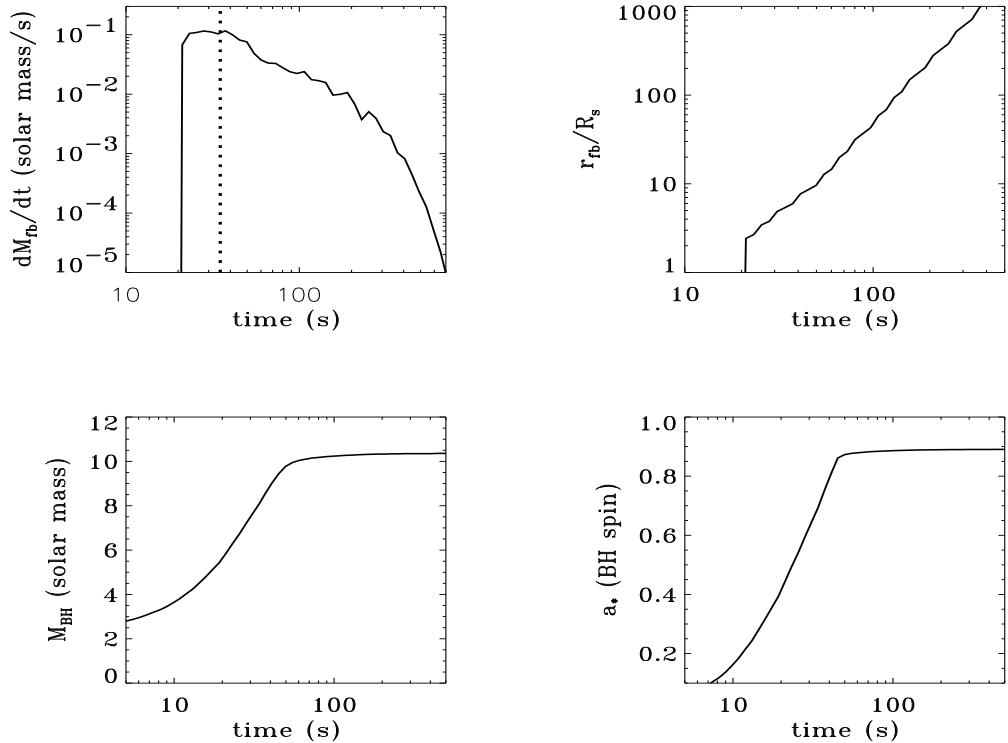}
\caption{\small The top left panel shows the rate at which gas rains
  down on to the accretion disk -- integrated over radius -- as a
  function of time; model 16T1 of Woosley \& Heger (2006) with the
  rotation profile shown in Fig. 2 was used for this calculation.  The
  top right panel shows the variation with time of the mean radius of
  the circular orbit on which most of the fall-back gas lands;
  the distance is in units of the Schwarzschild radius of
  the central black hole (the black-hole mass at 23s -- from direct
  collapse of the core -- is 5.7$M_\odot$).  The steep drop-off in
  fall-back rate at $\sim300$s is the result of a steep decrease of
  density with $r$ for the stellar model of Woosley \& Heger (2006) at
  $r\sim2.9\times10^{10}$cm (see Fig. 2). The lower left and right
  panels show the evolution of black hole mass and spin.}
\label{fig:n12}
\end{center}
\end{figure}

\subsection{Analytic Scalings}
It is instructive to try to understand the results shown in Fig. 3
using rough analytical estimates. The insights provided would be
useful for determining the mass fall-back rate for other rotation and
density profiles. We begin by expanding $t_{\rm eq}$ (eq. \ref{teq})
to second order in $\Omega/\Omega_k$:
\begin{equation}
t_{\rm eq} \approx t_s(r) + {\pi\over 2^{3/2}\Omega_k} \left[ 1 + {3\over4}
   \left({\Omega\sin\theta\over\Omega_k}\right)^2\right].
\label{teq1}
\end{equation}
The difference, $\delta r$, between the polar and equatorial radii of an
equal-collapse-time surface, i.e., the difference in $r$, for a fixed
$t_{\rm eq}$, between particles that start at $\theta=0$ and those
that start at $\theta=\pi/2$, is
\begin{equation}
\delta r \approx {3\pi\over 2^{7/2}} {[\Omega(r)/\Omega_k(r)]^2\over
    \Omega_k(r) d(t_s + \pi 2^{-3/2}\Omega_k^{-1})/dr} \sim
    {H_t\over 2} \left[{\Omega(r)\over\Omega_k(r)}\right]^2,
\end{equation}
where
\begin{equation}
H_t^{-1} \equiv \left| {d\over dr}\ln[t_s + \pi 2^{-3/2}\Omega_k^{-1}] \right|.
\end{equation}
Here we have assumed $t_s\sim \Omega_k^{-1}$, which is valid outside
of a small core region.  Since we see that $\delta r \ll r$ we may
treat the equal-collapse-time surface as spherical, and thus the mass
fall-back rate on to the disk is
\begin{equation}
\dot M_{\rm fb}(t) \equiv \pi\int dr\, r \dot \Sigma(r,t) \sim
    {d M(r)\over dr} {dr\over d t_{\rm eq}} \sim {4\pi r^2\rho(r) H_t\over t_{\rm eq}}
   \sim {4\pi r^2\rho(r)\over t_{\rm eq} \left| d\ln\Omega_k/dr \right|}.
\label{mfb1}
\end{equation} 

This simple analytical formula for the fall-back rate agrees to within
a factor of two with the numerical result shown in Figure 3. The
formula shows that $\dot M_{\rm fb}(t)$ is insensitive to the rotation
profile in the star. However, the fall-back radius where the stellar
matter circularizes has a strong dependence on $\Omega$, and is given
by
\begin{equation}
r_{\rm fb} \approx r \left({\Omega\over\Omega_k}\right)^2.
\label{rfb1}
\end{equation}
This is the mass-weighted average fall-back radius for a spherical
shell of gas; it is obtained from equations (\ref{ae}) \& (\ref{req}),
and the time dependence of $r_{\rm fb}$ is determined by the relation
between $r$ and $t_{\rm eq}$ given in equation (\ref{teq1}).

A good fraction of the gas in the core of the collapsing star will
have insufficient centrifugal support and will directly form a central
black hole.  Specifically, any material that circularizes inside the
innermost stable circular orbit (ISCO) of the black hole will fall
into the hole on a dynamical time.  The mass and dimensionless spin
$a_*$ of the initial black hole thus formed are given by the following
conditions:
\begin{equation}
R_{\rm isco}(M_r,a_*) = r {\Omega^2(r)\over\Omega_k^2(r)},
\qquad\qquad a_* = {cJ_r\over GM_r^2},
\end{equation}
where $M_r$ is the mass inside radius $r$ in the pre-collapse star,
$J_r$ is the angular momentum of this mass, and $R_{\rm isco}(M,a_*)$
is the radius of the ISCO for a black hole of mass $M$ and spin
parameter $a_*$ (Bardeen et al. 1972):
\begin{equation}
R_{\rm isco}(M,a_*) = {G M\over c^2} \left\{ 3 + z_2 - \left[(3-z_1)
    (3+z_1+2z_2)\right]^{1/2} \right\},
\end{equation}

\begin{equation}
z_1 = 1 + \left(1 - a_*^2\right)^{1/3}\left[ (1+a_*)^{1/3} + (1-a_*)^{1/3}
   \right], \quad{\rm and} \quad z_2 = \left(3 a_*^2 + z_1^2\right)^{1/2}.
\end{equation}
 We solve these equations numerically to calculate
the critical radius $r$ inside which all the mass falls directly into
the black hole, and thereby obtain the initial mass of the black hole
$M_{\rm BH}$ and its initial spin $a_*$.  For the model shown in
Fig. 2, we obtain $M_{\rm BH}=5.7 M_\odot$ and $a_* = 0.45$.

The rest of the material will have sufficient angular momentum to go
into orbit around the black hole.  Let us consider a model in which
the pre-collapse density profile is of the form $\rho(r)\propto
r^\tau$.  The index $\tau\equiv d\ln\rho/ d\ln r$ is shown
in Fig. 2 for the fiducial 14 $M_\odot$ stellar
model. The mass $M_r$ enclosed within $r$ increases with radius as
$r^{\tau'}$, where $\tau'=3+\tau$ for $\tau> -3$ and $\tau'\approx0$
when $\tau<-3$. Moreover, $H_t\approx|d\ln\Omega_k/dr|=
(3-\tau')/2r\approx1/r$, and $t_{\rm eq}\sim\Omega_k^{-1}\propto
r^{(3-\tau')/2}$.  Using these scalings we can calculate the time
dependence of the stellar mass fall-back rate on to the disk (using
eq. \ref{mfb1}) and the effective fall-back radius (eq. \ref{rfb1}):
\begin{equation}
\dot M_{\rm fb} \propto {r^{3+\tau}\over t_{\rm eq}} \propto t_{\rm eq}^{ (2\tau
    +\tau'+3)/(3-\tau')}, \quad\quad r_{\rm fb}\propto \Omega^2 r^{4-\tau'}
    \propto t_{\rm eq}^{ 2(4-\tau')/(3-\tau')} \Omega^2 \sim 
   t_{\rm eq}^{8/3} \Omega^2,
\label{mfb3}
\end{equation}
where the final expression corresponds to $\tau'=0$.

We see from Fig. 2 that $\tau\sim -2.5$ throughout most of the star
except near the surface, in the outermost 0.5$M_\odot$ layer, where
$\tau<-5$. Therefore, we expect $\dot M_{\rm fb}$ to decline approximately
as $t^{-0.5}$ while the interior of the star is collapsing, and $\dot
M_{\rm fb} \propto t^{-3}$ (or steeper) when the last $\sim0.5
M_\odot$ of the star near the surface is accreted; the steep fall-off
of $\dot M_{\rm fb}$ starts at about 100s after the beginning of the
stellar collapse. MacFadyen et al. (2001) find $\dot M_{\rm fb}
\propto t^{-2.4}$ on a time scale of 100--400s (see their Fig. 5)
when a shell at a distance of $\sim3\times10^{10}$ cm falls to the
center; the density profile in this shell corresponds to $\tau\sim-6$
(Fig. 2 in MacFadyen et al.)  and therefore their numerical scaling of
$\dot M_{\rm fb}$ is roughly in agreement with our crude analytical
scaling $\dot M_{\rm fb}\propto t^{-3}$.

The 1-D simulation of MacFadyen et al. (2001) has a strong forward
shock that controls the fall-back of stellar material on to the
central object and gives rise to $\dot M_{\rm fb}\propto t^{-1.7}$ for
$t\gta 400s$; this situation is unlikely to apply to the 3-D collapse
to a black hole. Unfortunately, there are no 3-D hydrodynamical or MHD
simulation results published that have time coverage $\gta\ 10^2$ s
for us to check our analytical scaling. The 1-D SNa simulation of 
Zhang, Woosley, \& Heger (2007) for a 25 $M_\odot$ star of radius 
$\sim 10^{12}$cm finds that the accretion rate is nearly constant for 
about 10$^3$s. This is most likely results from the accretion of 
the extended envelope of the progenitor star as our analytical
calculation suggests. We note that Zhang et al. (2007) do not see a 
phase at early times ($t\lta300$s) when the accretion undergoes
a sharp decline as suggested by GRB observations.

\section{Mass Accretion Rate and Jet Luminosity}

Figure 2 shows that there are two distinct zones in the pre-collapse
star: (i) the inner part of the stellar core (to the left of the
vertical dotted line) which collapses directly to form a black hole,
(ii) the outer part of the stellar core which goes into orbit around
the black hole and then accretes viscously.  In the following, we
explore a scenario in which accretion of material in the inner part of
zone ii produces the prompt GRB, while accretion of material near the
surface of the star, where the density decreases rapidly with $r$,
gives rise to the steeply declining flux at the end of the
$\gamma$-ray prompt emission.

\subsection{Approximate Prescription for the Accretion Rate}

To go from the mass fall-back rate $\dot{M}_{\rm fb}$ which we
estimated in the previous section to the mass accretion rate on the
black hole $\dot{M}_{\rm BH}$, we need to estimate (i) what fraction
of the gas that falls on the accretion disk actually makes it to the
black hole, and (ii) the time it takes the gas to accrete.  Both
issues have been discussed previously in the literature.

Narayan et al. (2001) showed that accretion in a fall-back disk can
occur via two distinct modes.  For high accretion rates and at small
radii, neutrino emission and cooling is efficient and accretion occurs
via a neutrino-dominated accretion flow (NDAF; Popham et al. 1999).
In this mode of accretion, essentially all the mass accretes on to the
black hole.  However, for lower fall-back rates and/or at larger
radii, the accretion is radiatively inefficient.  We then have an
advection-dominated accretion flow (ADAF; Narayan \& Yi 1994), and
only a fraction of the gas reaches the hole.  A number of later
investigators have studied the physics of these two regimes of
accretion in the context of GRB models (Kohri \& Mineshige 2002; Di
Matteo, Perna \& Narayan 2002; Lee, Ramirez-Ruiz \& Page 2004, 2005;
Janiuk et al. 2004, 2007; Kohri, Narayan \& Piran 2005; and references
therein).

Kohri et al. (2005) present detailed estimates of the advection
parameter $f_{\rm adv}$, viz., the fraction of the energy dissipated
in the disk that is advected with the gas, as a function of the local
accretion rate and the radius.  We will take $f_{\rm adv} = 0.5$,
which corresponds to half the energy being radiated 
and half being advected, as the
approximate boundary between the NDAF and ADAF regimes.  From Fig. 3
of Kohri et al. (2005), we find for the parameters of interest to us
that the boundary is located roughly at
\begin{equation}
\log\dot m \approx \log \left({r\over R_s}\right) -2.5,
\qquad \dot m \equiv {\dot M\over M_\odot{\rm s^{-1}}},
\qquad R_s \equiv {2GM\over c^2},
\label{NDAFADAF}
\end{equation}
where $r$ is the local radius, and $R_s$ is the Schwarzschild radius.

According to the prescription (\ref{NDAFADAF}), when the logarithm of
the mass accretion rate $\dot m$ (in solar masses per
second) at the outer perimeter of the disk at radius $r_{\rm d}$ is greater 
than $\log (r_d/R_s)-2.5$, the accretion occurs via an NDAF and all
the fall-back material accretes on to the black hole.  However, when
$\log \dot m$ is smaller than this limit, accretion occurs
at least partially via an ADAF.

An important feature of an ADAF is that it generally has a strong mass
outflow (Stone, Pringle \& Begelman 1999; Igumenshchev \& Abramowicz
2000), which is believed to be driven by a positive Bernouilli constant
(Narayan \& Yi 1994, 1995).  Because of this, at each radius a fraction of
the accreting mass is lost in a wind, and so the net accretion rate
decreases as we go to smaller radii.  The exact functional form of
this decrease is uncertain.  In the most extreme case, we expect to
have a convection-dominated accretion flow (CDAF, cf. Narayan,
Igumenshchev \& Abramowicz 2000; Quataert \& Gruzinov 2000; Narayan et
al. 2001), in which the accretion rate decreases linearly with $r$.
More generally, we expect a scaling of the form
\begin{equation}
\dot{m}(r) \approx \dot m_{\rm acc}\left({r\over r_d}\right)^s, 
\qquad 0 \leq s \leq 1,
\label{mdot3}
\end{equation}
where $\dot m_{\rm acc}$ is the accretion rate in units of solar
mass per second at the outer boundary of the disk, and $\dot m(r)$
is accretion rate ($M_\odot$ s$^{-1}$ unit) at radius $r$.
Pen, Matzner \& Wong (2003) estimated $s\sim0.8$ from 3D MHD
simulations, while Yuan, Quataert \& Narayan (2003) deduced $s\sim0.3$
from modeling the radiatively inefficient accretion flow in the
Galactic Center source Sgr A$^*$.  As an added complication, $s$
probably depends on the degree of advection (i.e., the value of
$f_{\rm adv}$).  There is therefore considerable uncertainty regarding
the value of $s$, though a choice of $s\sim0.5$ is probably
reasonable.  In the following, we leave $s$ as a free parameter.  Thus
we take $\dot{m}$ to decrease as $r^s$ so long as accretion occurs via
an ADAF ($\log\dot m$ less than the limit in eq. \ref{NDAFADAF}) and
to be independent of $r$ once the accretion flow becomes an NDAF
(larger values of $\log\dot m$).  We also assume that all the mass
that reaches the ISCO falls into the black hole.  We then have three
different regimes of accretion, each with its own prescription for the
mass accretion rate:
\begin{eqnarray}
I:\qquad \log\mbh &=& \log\macc, \qquad {\rm if}~\log\left({\rd\over
R_s}\right)-2.5 \leq \log\macc,\label{NDAF} \\
II:\qquad \log\mbh &=& {1\over(1-s)}\left[\log\macc-s\log\left({\rd\over R_s}
\right)+2.5\,s\right], \nonumber\\
&~&{\rm if}~\log\left({R_{\rm isco}\over
R_s}\right)+ s \log\left({\rd\over R_{\rm isco}}\right)-2.5 \leq \log\macc <
\log\left({\rd\over
R_s}\right)-2.5, \label{NADAF}\\
III.\qquad \log\mbh &=& \log\macc - s \log\left({\rd\over R_{\rm isco}}\right), 
\nonumber\\
&~&{\rm if}~\log\macc < \log\left({R_{\rm isco}\over
R_s}\right)+ s \log\left({\rd\over R_{\rm isco}}\right)-2.5,\label{ADAF}
\end{eqnarray}
The three regimes correspond to: (I) pure NDAF (eq. \ref{NDAF}), (II)
ADAF on the outside and NDAF on the inside (eq. \ref{NADAF}), and
(III) pure ADAF (eq. \ref{ADAF}).  In these expressions, $R_s$ is the
Schwarzschild radius and $R_{\rm isco}$ is the radius of the ISCO
corresponding to the current mass and spin of the black hole.

The disk radius $r_d$ and the outer mass accretion rate $\macc$ are
determined by the current disk mass and angular momentum. Let $M_d(t)$
be the mass of the disk at time $t$ and $J_d(t)$ the total angular
momentum in the disk. The effective radius of the disk $r_d$ is then
defined by
\begin{equation}
{J_d\over M_d} = j(r_d) = \left(GM_{\rm BH} r_d\right)^{1/2},
\label{am1}
\end{equation}
where $j(r)$ is the (Newtonian) specific angular momentum of an
orbiting particle at radius $r$. The mean rate at which mass empties from 
the disk as a result of accretion is
\begin{equation}
\dot{M}_{\rm acc} = {M_d\over t_{\rm acc}},
\label{macc}
\end{equation}
where the accretion time scale $t_{\rm acc}$ for a thick disk of vertical
scale height $\sim r_d$ is approximately given in
terms of the kinematic coefficient of viscosity $\nu$ by
\begin{equation}
t_{\rm acc} \sim {r_d^2\over\nu(r_d)} = \left({v_K\over c_s}
\right)^2 {1\over\alpha\Omega_k} \sim {2\over \alpha\Omega_k}.
\label{tacc}
\end{equation}
Here, $\alpha$ is the standard dimensional viscosity parameter
(Shakura \& Sunyaev 1973), which has a value $\sim0.01-0.1$, and the
factor of 2 in the final expression is approximately correct for a
fully radiatively inefficient accretion flow.

The mass and angular momentum of the disk change with time; they
increase as a result of fall-back from the stellar envelope and they
decrease as a result of accretion.  Thus we may write
\begin{eqnarray}
\dot{M}_d &=& \dot{M}_{\rm fb} - \dot{M}_{\rm acc}, \label{mdt0}\\
\dot{J}_d &=& \dot{J}_{\rm fb} - \dot{J}_{\rm acc}.
\label{jdot}
\end{eqnarray}
The fall-back model described in \S~2 gives the fall-back terms
$\dot{M}_{\rm fb}$ and $\dot{J}_{\rm fb}$, while equation (\ref{macc})
gives $\dot{M}_{\rm acc}$. The angular momentum loss rate,
$\dot{J}_{\rm acc}$, due to accretion results from mass falling into
the BH -- $j(R_{\rm isco}) \dot{M}_{\rm BH}$ -- plus angular momentum
carried away by the wind. We assume that the specific angular momentum
in the wind is equal to that of the gas in the disk at the radius
from which the wind originates. It is then straightforward to integrate
over radius and calculate the net angular momentum loss in the wind.
Adding the two contributions, we have
\begin{equation}
\dot{J}_{\rm acc} = j(R_{\rm isco}) \dot{M}_{\rm BH} +{2s\over
(2s+1)} j(r_d)\dot{M}_{\rm acc}\left[ 1 - \left({r_t\over r_d}
      \right)^{{(2s+1)/2}}\right],
\label{jacc}
\end{equation}
where
\begin{equation}
r_t = R_s \left[10^{2.5} \dot{m}_{\rm acc} (r_d/R_s)^{-s}\right]^{{1\over 1-s}},
\label{rt2}
\end{equation}
is the radius where a transition (if any) from NDAF to ADAF occurs; $\dot{m}_{\rm acc} \equiv \dot{M}_{\rm acc}/(1 M_\odot$ s$^{-1}$).

The rate of increase of the black hole mass and angular momentum are:
\begin{equation}
{dM_{\rm BH}\over dt} = \dot{M}_{\rm BH}, \qquad\qquad
{dJ_{\rm BH}\over dt} = \dot{M}_{\rm BH} \,j_{\rm isco},
\label{mdotbh}
\end{equation}
where $j_{\rm isco}$ is the specific angular momentum of a particle
on a circular orbit at the ISCO (see Bardeen et al. 1972):
\begin{equation}
j_{\rm isco} = \left(G M_{\rm BH} R_{\rm isco}\right)^{1/2} { R_{\rm isco}^2 - a_* R_s 
      \left(R_{\rm isco} R_s/2\right)^{1/2} + a_*^2 R_s^2/4\over R_{\rm isco}
     \left[ R_{\rm isco}^2 - 3R_{\rm isco} R_s/2 + a_* R_s \left(R_{\rm isco} R_s/2
     \right)^{1/2} \right]^{1/2}  }.
\end{equation}

We solve equations (\ref{am1})--(\ref{mdotbh}) numerically and
determine the accretion rate on to the BH and the evolution of the
accretion disk.  However, when $\dot{M}_{\rm fb}=0$, or when $t_{\rm acc}\ll
(r_d^3/G M_{\rm BH})^{1/2}$, these coupled equations can be solved
analytically as described in the next subsection.

\subsection{Analytical solutions}

A formal solution of equation (\ref{mdt0}) can be shown to be-
\begin{equation}
M_d(t) = M_d(t_0) \exp\left(-\int_{t_0}^t dt_1\, t_{\rm acc}^{-1}\right) 
    + \int_{t_0}^t dt_1\, \dot{M}_{\rm fb}(t_1) \exp\left(-\int_{t_1}^t dt_2 
\, t_{\rm acc}^{-1}\right).
\label{Mdt1}
\end{equation}
For $t_{\rm acc}\ll t$, i.e., rapid accretion, the first term on the right
is very small and can be neglected. In this limit, the disk mass and
the accretion rate are determined by the instantaneous value of the
mass-fall-back rate, i.e. $\dot M_{\rm acc} =M_d/t_{acc} \approx
\dot{M}_{\rm fb}(t)$, and the jet power (see \S~3.3) tracks the
fall-back rate. This result comes in handy when we wish to understand
various features in the early x-ray lightcurve of GRBs and their
relationship to the structure of the progenitor star.

Another special case of considerable interest is where the stellar
collapse leaves behind a reservoir of gas in a disk around the black
hole, and no further mass is being added, i.e. $\dot{M}_{\rm fb}=0$.
The accretion of gas from the reservoir on to the BH can keep the
relativistic jet going for some period of time. We now estimate this
accretion rate.

Equations (\ref{macc})--(\ref{jacc}) can be combined when $\dot{M}_{\rm fb}=0$
to obtain
\begin{equation}
\dot{M}_d = - {\alpha G^2 M_d^4 M_{\rm BH}^2 \over J_d^3},
\label{dotMd1}
\end{equation}
and
\begin{equation}
\dot{J}_d = - {2s\alpha\over 2s + 1} {G^2 M_d^3 M_{\rm BH}^2 \over J_d^2}.
\label{dotJd1}
\end{equation}
In deriving equation (\ref{dotJd1}) we assumed $r_t\ll r_d$, and we
neglected the angular momentum deposited on to the BH in comparison to
that carried away by the wind.

Equations (\ref{dotMd1}) and (\ref{dotJd1}) can be easily solved to yield
\begin{equation}
{ J_d(t)\over J_d(t_0) } = \left[ { 
 M_d(t)\over M_d(t_0) } \right]^{2s/(2s+1)}
  \quad\quad{\rm and} \quad\quad r_d\propto (J_d/ M_d)^2
   \propto M_d^{ -2 /(2s+1)} \,. 
\label{Jdt}
\end{equation}
We substitute this solution back into equation (\ref{dotMd1}) to
eliminate $J_d$, and solve the resulting equation to find the
accretion rate at $r=r_d$:
\begin{equation}
{\dot{M}_d(t)\over M_d(t_0) } = {1\over t'_{\rm acc}} \left[ 1 +
     {3\over (2s+1)}{(t-t_0)\over t'_{\rm acc} } \right]^{-(2s+4)/3},
\label{Mdt}
\end{equation}
where
\begin{equation}
t'_{\rm acc} \equiv {2\over \alpha G^2 M_{\rm BH}^2 } \left[ { J_d(t_0)\over M_d(t_0) }
   \right]^3 = t_{\rm acc}(t_0).
\label{taccp}
\end{equation}
In the limit when $s=0$, the accretion rate declines with time as
$t^{-4/3}$, consistent with the similarity solution described by
Ogilvie (1999).  The decline is faster for larger values of $s$, with
the fastest decline being $t^{-2}$ for $s=1$.

Combining equations (\ref{Jdt}) \& (\ref{Mdt}) we obtain
\begin{eqnarray}
{ J_d(t)\over M_d(t) } &=&  { J_d(t_0)\over M_d(t_0) } 
     \left[ 1 + {3\over 2s+1}{(t-t_0)\over t'_{\rm acc} } \right]^{{1/3}}, \\
\label{JdMd}
r_d(t) &=& {1\over GM_{\rm BH}} \left[ { J_d(t)\over M_d(t) } \right]^2
 = r_d(t_0) \left[ 1 + {3\over 2s+1}{(t-t_0)\over t'_{\rm acc} } \right]^{
   {2/3}}, \\
\label{rdt}
\Omega_k(t) &=& \sqrt{ {GM_{\rm BH}\over r_d^3}}
   = \Omega_k(t_0) \left[ 1 + {3\over 2s+1}{(t-t_0)\over t'_{\rm acc} } 
     \right]^{-1} \propto t^{-1}.
\label{OmKt}
\end{eqnarray}
The accretion time, $t_{\rm acc}\approx 2/(\alpha \Omega_k)$, 
asymptotically approaches $3t/(2s+1)$ for $(t-t_0)/t'_{\rm acc}\gg1$.

The accretion rate on to the BH is $\sim |\dot{M}_d| (r_d/R_s)^{-s}$:
\begin{equation}
\dot{M}_{\rm BH}(t) \approx \dot{M}_{\rm BH}(t_0) \left[ 1 + 
    {3\over 2s+1}{(t-t_0)\over t'_{\rm acc} }\right]^{-{4(s+1)/3}}.
\label{mdotbh1}
\end{equation}
The rate declines as $t^{-4/3}$ for $s=0$, but much more steeply as
$t^{-8/3}$ when $s=1$; it goes as $t^{-2}$ for the intermediate value
$s=0.5$.

\subsection{Prescription for the Jet Luminosity}

To convert the mass accretion rate $\dot M_{\rm BH}$ to the power
output $L_{\rm jet}$ in a relativistic jet, we need to estimate the jet
efficiency factor $\eta_j$:
\begin{equation}
L_{\rm jet} = \eta_j\dot{M}_{\rm BH} c^2.
\label{Lj}
\end{equation}
Despite many years of study, the details of how relativistic jets are
launched from accreting black holes are still poorly understood.  The
efficiency factor $\eta_j$ is likely to depend on many details, but it
is probably most sensitive to the spin of the black hole.  In this
paper, we make use of the following approximate prescription obtained
by McKinney (2005) by fitting numerical results from  GRMHD simulations:
\begin{equation}
\eta_j \approx 0.07 \left({a_*\over1+\sqrt{1-a_*^2}}\right)^5.
\label{etaj}
\end{equation}
According to this prescription, the efficiency is a very steeply
increasing function of the black hole spin: $\eta_j\sim10^{-4}$ for
$a_*\sim0.5$, $\eta_j\sim10^{-3}$ for $a_*\sim0.75$,
$\eta_j\sim10^{-2}$ for $a_*\ \gta\ 0.9$.

We note that McKinney's simulations corresponded to a non-radiating
accretion flow, i.e., an ADAF.  Thus, for our problem, the above
prescription is valid only for the ADAF phase of accretion; it is not
clear what we should do when accretion occurs via an NDAF.  For
simplicity, we use the same prescription for the NDAF phase as well.
Fortunately, most of the accretion in the fall-back disk occurs via an
ADAF, so the error is probably not serious.

Given a model for the density and rotation profile of the pre-collapse
star, the prescriptions given in \S\S~3.1-3.3 allow us to estimate the
jet luminosity $L_{\rm jet}$ as a function of time.  We are thus ready to
consider the implications of this model for GRBs.

\subsection{Prompt GRB Emission and Rapid Shutoff}

We begin by discussing the first problem of interest to us, viz., the
origin of the prompt GRB emission and the reason for the abrupt
shutoff of this emission.  As already mentioned, we associate the
prompt GRB with accretion of the outer regions of the stellar core
($M>8M_\odot$ in Fig. 2).  Using the $14M_\odot$ stellar model shown
in Fig. 2, we calculated the mass accretion rate via equations
(\ref{NDAF})--(\ref{ADAF}) for different values of the parameter $s$.
We then computed the corresponding jet luminosity as a function of
time using equation (\ref{Lj}).  The results are shown in Fig. 4.
Also shown is the powerlaw index for the temporal decline of the jet
power.

For an initial accretion rate of $\sim 10^{-1} M_{\odot}$ s$^{-1}$
(Fig. 3) and $\eta_j\sim 10^{-2}$, we expect the jet power to be $\sim
10^{51}$ erg s$^{-1}$, as seen in the numerical results shown in
Fig. 4.  This is roughly consistent with the power observed in
long-GRBs.  Thus, at least in terms of the overall energetics, the
model gives fairly reasonable results.

\begin{figure}[h!]
\begin{center}
\includegraphics[height=3.9in,width=5.95in]{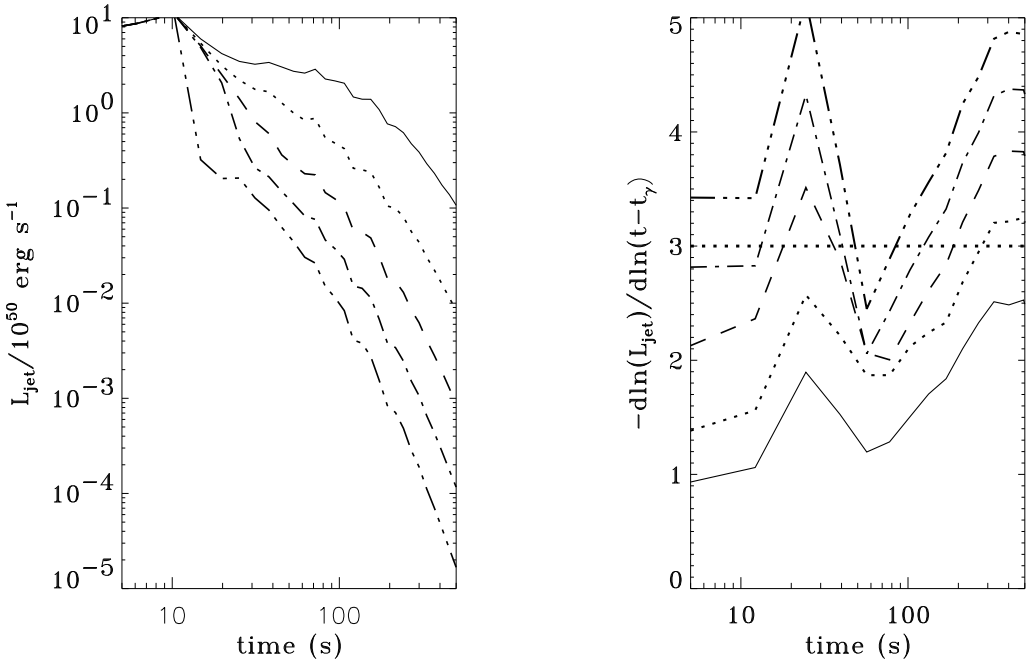}
\caption{\small {\bf Left panel:} shows the jet power $L_{\rm jet}$ (this
  is the total jet power, not the isotropic equivalent luminosity) as
  a function of time for the 14$M_\odot$ pre-collapse stellar model
  and the rotation profile $\Omega(r)$ shown in Fig. 2. The time axis has been shifted by 33s, 
  so that t=0 here is approximately when we see the start of the gamma-ray 
  burst; a relativistic jet produced during the initial $\sim10$s (in the rest
  frame of an observer at the center of the star) of high accretion rate
  clears a polar cavity in the star and a short time thereafter, in
  the observer frame, $\gamma$-rays are generated. The five different
  curves correspond to five different values of the parameter $s$ (see
  eq. \ref{mdot3} for definition); from top to bottom, the curves
  correspond to $s$=0 (solid line), 0.25, 0.5, 0.75 \& 1.0;
  $\alpha=0.1$ for all of these curves. A smaller fraction of gas
  reaches the black hole for larger $s$ (the remainder leaves the
  system via a wind from the disk). This causes the overall luminosity
  to be lower and also the power to decline more rapidly. The steep dropoff
  of jet power at $t\sim 10$s, for $s\gta 0.5$, occurs when the accretion
  flow makes a transition from an NDAF in which all the gas reaches the
  black hole to an ADAF in which only
  a small fraction of the in-falling gas reaches the black hole.
  {\bf Right panel:} Shows -$d\ln(L_{\rm jet})/d\ln(t-t_\gamma)$,
  where $t_\gamma$=33s. The different curves here correspond to different
  values of $s$, exactly as in the left panel.  For
  -$d\ln(L_{\rm jet})/d\ln(t-t_\gamma)\ \gta\ 3$ the observed lightcurve
  will be dominated by off-axis emission and will decline as
  $t^{-2-\beta}\sim t^{-3}$.  }
\label{fig4}
\end{center}
\end{figure}

The abrupt decline of the prompt emission after a period of activity
is more challenging.  The fastest possible decline is limited by the
curvature of the $\gamma$-ray source surface and is given by
$t^{-2-\beta}$ (Kumar \& Panaitescu, 2000; $\beta$ is the spectral
index of the radiation, i.e., $f_\nu\propto\nu^{-\beta}$) when the jet
opening angle is larger than the inverse of the jet Lorentz factor.
Declines of this order have been observed with the x-ray telescope
aboard Swift (e.g., Tagliaferri et al. 2005; O'Brien et al. 2006).
Such a rapid rate of decline is possible only if the intrinsic jet
power itself declines faster than $t^{-2-\beta}\sim t^{-3}$.  Thus, in
order to explain the observed steep decline, we require the jet power
in our model to satisfy $-d\ln L_{\rm jet}/d\ln(t-t_\gamma) > 3$,
where $t_\gamma$ is the time when gamma-ray emission is first observed.

Figure 4 shows that at least some of our models do satisfy this
requirement.  Specifically, we find that the decline is faster than
$t^{-3}$ after about 100 s whenever $s\ \gta\ 0.5$. The rapid decline
results from the steeply falling density profile in the outer part of
the GRB progenitor star (see Fig. 2), coupled with the fact that a
progressively decreasing fraction of the fall-back mass reaches the
black hole for larger values of $s$.  The numerical results are
consistent with the analytical scalings of \S\S~2.4, 3.2; $\L_{\rm jet}
\propto \dot M_{\rm fb} r_{\rm fb}^{-s}\propto t^{(2\tau+3-8s)/3}$ for
an ADAF. The results shown in Fig. 4 are for $\alpha=0.1$. The jet 
power is not very sensitive to $\alpha$ because as long as $t_{\rm acc}\propto
\alpha^{-1}$ is less than $t$, $\dot M_{\rm acc}\approx \dot M_{\rm
  fb}$; for a larger $\alpha$, $t_{\rm acc}$ is smaller, and that
results in a slightly steeper decline of $L_{\rm jet}(t)$ since $\dot M_{\rm
  acc}$ is now equal to $\dot M_{\rm fb}$ averaged over a smaller time period.

The sharp drop in $L_{\rm jet}$ at $t\sim 10$s results from 
the transition from an NDAF 
to a fully ADAF solution; the sharp drop is because the transition radius ($r_t$)
 is a very steep function of $\dot M_{\rm acc}$ and $r_d$ for $s\ \gta\ 0.5$ (see 
eq. \ref{rt2}).

\begin{figure}[h!]
\begin{center}
\includegraphics[height=3.9in,width=5.95in]{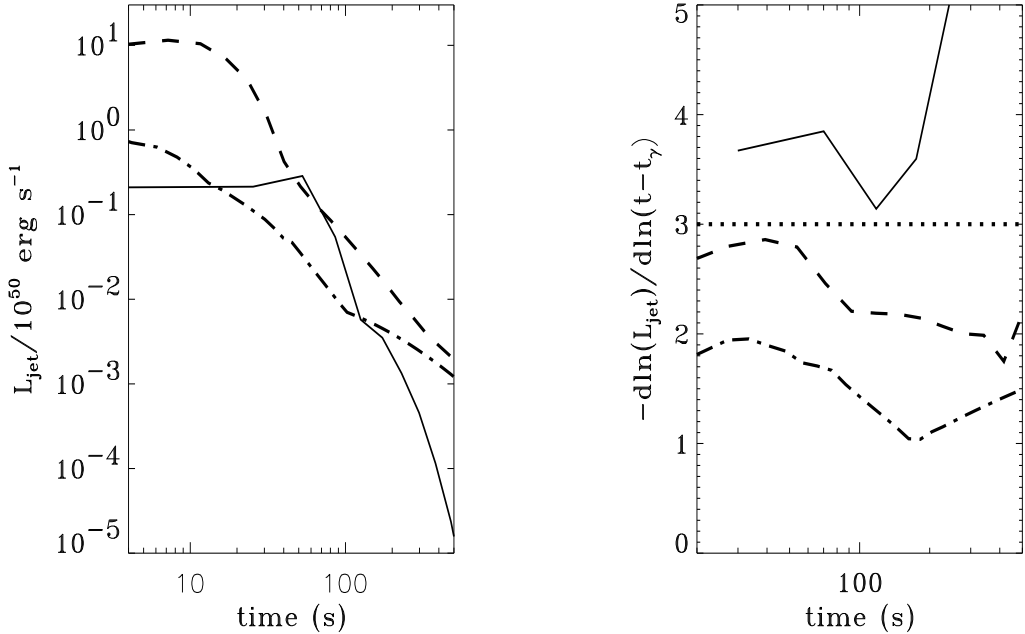}
\caption{\small Shows the effect of changing the rotation rate in the
  stellar core on the jet power. All the models assume $s=0.5$,
  $\alpha=0.1$.  {\bf Left panel:} Shows the jet power as a function
  of time for the same stellar model as in Fig. 2, but with different
  rotation rates.  The solid curve is for a rotation rate equal to 1/3
  of $\Omega(r)$ shown in the lower right panel of Fig. 2. The other
  models correspond to a rotation rate of 3 times
  (dashed line) and 9 times (dot-dashed) $\Omega(r)$ in Fig. 2.
  When the rotation rate is about eight times smaller
  than in Fig. 2 the entire stellar core collapses directly to a
  black-hole without first forming a disk.  {\bf Right panel:} Shows
  -$d\ln(L_{\rm jet})/d\ln(t-t_\gamma)$ for the same three rotation profiles;
   the time $t_\gamma$ is when we see the first $\gamma$-ray
  photons from the burst and is roughly the time it takes for the
  relativistic jet to emerge at the stellar surface. We took
  $t_\gamma=110$s, 20s \& 12s for $\Omega=$ 1/3, 3 \& 9
  times the rate in Fig. 2.  }
\label{fig5}
\end{center}
\end{figure}

The duration of the steep decline is determined by the mass, radius,
and rotation rate of the progenitor star's core; the mass and radius
set the collapse time-scale $2(R^3/GM)^{1/2}$, and the rotation rate
determines the fraction of the core that collapses directly to a
black hole.  The effect of core rotation on the jet luminosity is shown
in Figure 5. The peak jet power is much smaller when $\Omega$ in the core
is either much larger, or much smaller, than the value shown in fig. 2.
For large $\Omega$, the accretion disk radius, and $t_{acc}$, are larger
and hence the jet power is smaller; much of the stellar mass in this case
is ejected as a sub-relativistic, bi-polar, wind launched from the disk. 
Whereas for small $\Omega$, a larger fraction of the core collapses directly 
to form a BH, and the total mass of gas available to form a disk and power 
the jet is smaller. When the core rotation rate is
about eight times smaller than the value shown in Fig. 2, the angular
momentum is insufficient to form a disk and the entire core collapses
to a black hole.  In this case there would be no GRB. A variation in
core rotation rate by a factor of 27 -- the range considered in Fig. 5
-- leads to a change in the peak jet power by a factor $\sim 20$, and
the duration over which the luminosity is high varies from $\sim 10$s
to 10$^2$s; all time scales are in the host-galaxy rest frame. The
duration of the steep decline phase on the other hand is $\sim 400$s
for different $s$ values (Fig. 4) and for rotation speeds $\gta$ the
model considered in Fig. 2. 

The jet luminosity declines as $\sim t^{-2}$ for larger $\Omega$ (see
Fig. 5) due to the fact that $t_{\rm acc}$ becomes greater than 
$t$,\footnote{The fall-back radius $r_{\rm fb} \propto \Omega^2$
  (eq. \ref{rfb1}), so $t_{\rm acc} \propto \Omega_k(r_{\rm fb})^{-1}
  \propto \Omega^{3}$.} and this leads to the accretion rate on to the
BH declining as $t^{-4(s+1)/3}$ (eq. \ref{mdotbh1}) when $\dot{\rm
  M}_{\rm fb}$ decreases rapidly (during the period the outer part of the
star is collapsing).  As we have already noted, a decline of jet-power
as $\sim t^{-2}$ or slower is not consistent with early x-ray
observations of GRBs (eg. O'Brien et al. 2006).  This constraint
provides a limit on the rotation rate in the outer part of the
progenitor star in our model.

\subsection{Explanation for the Plateau in the X-ray Lightcurve}

We now consider the second major puzzle in Swift observations of GRBs,
viz., the presence of a plateau in the lightcurve for about $10^4$ s
in $\sim50\%$ of long-duration GRBs.  First, we would like to
understand what keeps the GRB engine operating for such a long time
after the shutoff of the prompt burst.  Second, we would like to
explain how the engine is able to maintain a very shallow decline of
luminosity $\sim t^{-0.5}$ during this entire time.  Third, we would
like to reproduce the sudden and sharp drop of luminosity at the end
of the plateau seen in several GRBs, e.g., 060413, 060607A, 070110
(Liang et al. 2007).  Needless to say, it is very difficult to do all
this within an accretion model.

We describe two possible solutions to the plateau problem.  One
solution (\S~3.5.1) invokes a very small viscosity parameter $\alpha$,
such that the viscous accretion time of the fall-back disk is
comparable to the duration of the plateau $\sim10^4$ s.  The second
solution (\S\S~3.5.2, 3.5.3) postulates continued fall-back of gas on
the reservoir/accretion disk for about $10^4$ s, which is much longer
than the free-fall time $\sim 500$ s from the surface of our model
star (Fig. 2).

\subsubsection{Plateau as a result of small $\alpha$}

Figure 6 shows the long term evolution of the jet luminosity $L_{\rm
  jet}$ for different values of the viscosity parameter $\alpha$. For
$\alpha=0.1$ (our standard value), we find that $L_{\rm jet}$ declines
rapidly with time for $t\ \gta\ 10^2$s. However, for
$\alpha\ \lta\ 10^{-2}$, $L_{\rm jet}(t)$ has a plateau starting at
about 200s, and the duration of the plateau increases with decreasing
$\alpha$.  These results are for the pre-collapse stellar model of
Woosley \& Heger (2006) shown in Fig. 2, which is compact ($R_*=$
$5\times10^{10}$cm) and has its density decreasing faster than $r^{-4}$
near the surface; the free-fall time at the surface of this model is
about 500s. 

\begin{figure}[h!]
\begin{center}
\includegraphics[height=5.8in,width=5.5in]{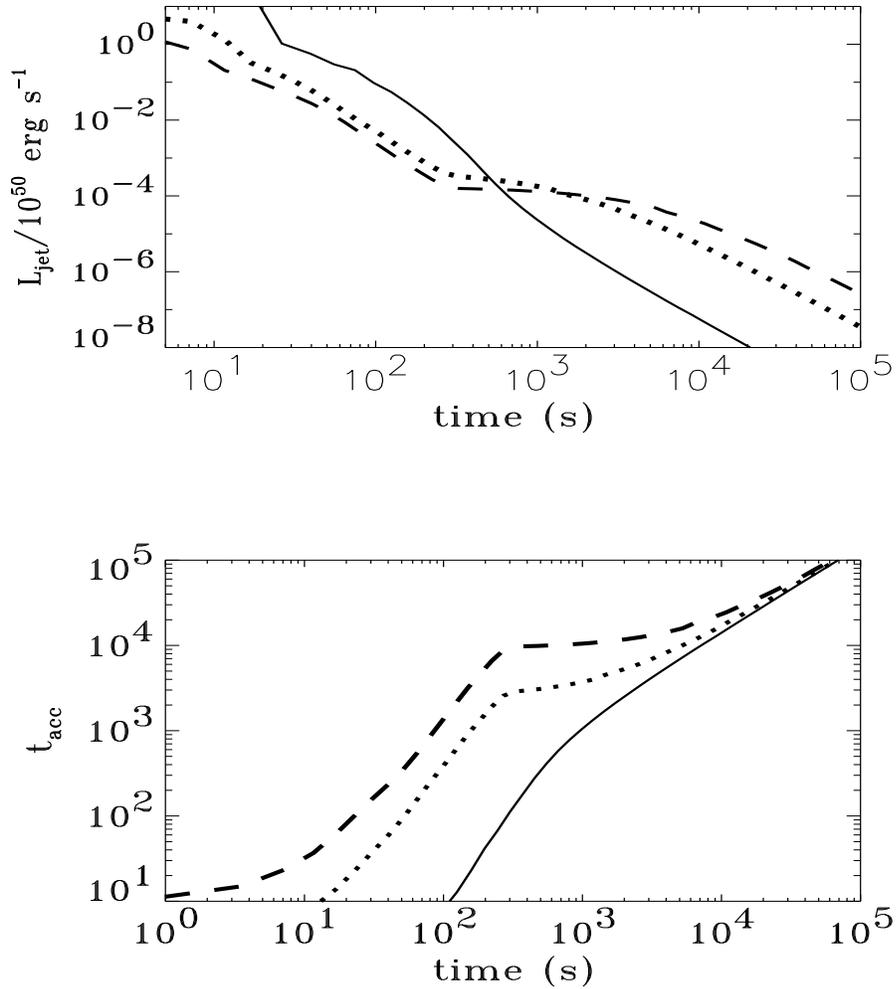}
\caption{\small Shows the long term behavior of the jet power $L_{\rm
    jet}$ (top panel) and accretion time scale $t_{\rm acc}$ (lower
  panel) for three choices of $\alpha$: 0.1 (solid line), 0.01 (dotted
  line), 0.001 (dashed line).  All the models correspond to $s=0.75$.
  For the $\alpha=0.1$ model, $\Omega(r)$ is as shown in Fig. 2, while
  for the other two models it was taken to be four times larger.  Note
  that the jet power shows a plateau for smaller values of $\alpha$.
}
\label{fig6}
\end{center}
\end{figure}

For $\alpha\ \gta\ 0.1$, the viscous accretion time is shorter than
the dynamical fall-back time (Fig. 6), and therefore the accretion
rate on to the BH tracks the rate at which mass is added to the disk
(see the analytical calculation in \S3.2 and the discussion following
eq. \ref{Mdt1}).  This leads to a rapid decrease of $L_{\rm jet}$ for
$t\ \lta\ 300$s, which nicely explains the sudden shutoff of the
prompt emission as described in \S~3.4.  It is, however, a problem for the
plateau.  For $t\ \gta\ 300$ s, when $\dot{M}_{\rm fb}=0$ (since the
entire star has collapsed), the power continues to fall quite rapidly
in this model.  From our previous analysis (see eq.  \ref{mdotbh1}),
we expect the jet power to decline as $t^{-4(s+1)/3}$, which is
consistent with the numerical result shown by the solid curve in
Fig. 6, and is too steep to produce a viable plateau.

For $\alpha\ \lta\ 10^{-2}$, the accretion time ($t_{\rm acc}$) is
longer than the dynamical time (Fig. 6), and this allows the disk mass
to build up to a substantial value ($M_d \sim 1.0 M_\odot$). In this
case when $\dot{M}_{\rm fb}$ drops sharply -- during the collapse of
the outer, low density, envelope -- the accretion rate $\dot{\rm
  M}_{\rm acc}$ and the disk radius remain nearly constant for a time
of order $t_{\rm acc}$ (see eqs. \ref{Mdt} \& \ref{rdt}).  As a
result, the jet-power remains nearly constant until $t\sim t_{\rm
  acc}$. For $t\ \gta\ t_{\rm acc}$, we switch back to the canonical
$L_{\rm jet}\propto t^{-4(s+1)/3}$ behavior as described above.

We should note that, even with the small values of $\alpha$ considered
here, a long-lasting plateau can be obtained only if $\Omega$ is
larger by a factor of about 5 than the model considered in Fig. 2.
The reason is that the fall-back radius $r_{\rm fb}\propto \Omega^2$ (eq.
\ref{mfb3}), and $t_{\rm acc}\propto r_{\rm fb}^{3/2}\propto\Omega^{3}$
(eq. \ref{tacc}).  Since we require a long viscous time $t_{\rm acc}$,
we must have a large $\Omega$.  Of course, a faster rotating star
would need to be more compact in order to be bound, and this partially
offsets the larger $t_{\rm acc}$ we would get for larger
$\Omega$. Nevertheless, a star rotating faster by a factor of 7 has
$t_{\rm acc}\gg t$ even for $\alpha = 0.01$ and will give a plateau in
$L_{\rm jet}$ lasting for $10^4$s.

On the whole, we believe this is a reasonable scenario to explain the
plateau in GRB lightcurves.  However, it has a few problems.  First,
in order to have a plateau extending up to $10^4$ s as seen in many
GRBs, we need to decrease $\alpha$ almost to $10^{-3}$.  This value is
much smaller than what is expected from the magneto-rotational
instability ($\alpha\sim0.01-0.1$; Stone et al. 1996).  Of course, GRB accretion
disks represent a very different regime (of density, temperature,
radiation processes) than the disks one usually encounters in
astrophysics, and it is possible that there is a reason why $\alpha$
might be unusually small in these disks.  Another problem is that the
models with the most pronounced plateaus have a shallower decline of
the lightcurve between the prompt burst and the plateau; we tend to
obtain $L_{\rm jet}\propto t^{-2.5}$ instead of $t^{-3.5}$, so these
models do a poor job of explaining the rapid shutoff of the prompt
burst (\S~3.4).  Another potential problem is that these models always
have $L_{\rm jet}$ declining as $t^{-4(s+1)/3}$ beyond the plateau.
While this is acceptable for many GRBs, there are several bursts for
which the luminosity drops much more rapidly at the end of the plateau.
The present scenario cannot explain these bursts.

In summary, a compact stellar progenitor (e.g., Woosley \& Heger 2006
model) combined with an accretion disk with $\alpha\ \lta\ 0.01$, can
provide a natural explanation for the small subset of GRBs that have
short-lived plateaus in their x-ray lightcurves, have flux declining
less rapidly than $t^{-3}$ before the plateau, and flux declining as
$\sim t^{-2}$ immediately after the plateau (Fig. 6, dotted line).
For the remaining more extreme bursts, we need another scenario.

\subsubsection{Plateau as a result of continued mass fall-back} 

We now return to our standard accretion model with $\alpha\sim0.1$ and
$t_{\rm acc} < t$.  In this case, in the absence of mass fall-back,
the jet power will decline steeply as $L_{\rm jet}\sim t^{-4(s+1)/3}$,
which is much too steep to explain the plateau.  The only way we can
avoid the luminosity drop is by having continued mass-fall-back for
the duration of the plateau $\sim10^4$ s. We discuss this solution
here and in the next subsection.

We describe the extended fall-back with three parameters: the amount
of mass involved in the fall-back which is about $1M_\odot$ for the
model calculations presented here, the time dependence of the mass
fall-back rate which we take to be a power-law with a specified index,
and the angular momentum of the fall-back gas $\dot{\rm
  J}_{\rm fb}/\dot{M}_{\rm fb}\equiv j$ which we assume to be independent
of time without restricting the solution space much.  Once the
fall-back is turned on (say at time $t_0$) the solution evolves
quickly so that $t_{\rm acc} \to 2j^3/(\alpha G^2 M_{\rm BH}^2)$ (see
eqs. \ref{Mdt} \& \ref{taccp}). Initially, it is possible for
$\dot{M}_{\rm BH}$ to fall rapidly, provided $\dot{M}_{\rm fb}(t_0) <
\dot{M}_{\rm acc}(t_0)$ and $t_{\rm acc}(t_0) < t_0$.  However, on
a timescale of max($t_0$, $t_{\rm acc}$), a quasi-steady state is
established with $\dot{M}_{\rm acc}(t)\sim \dot{M}_{\rm fb}(t)$
(see the discussion following eq. \ref{Mdt1}), and the disk radius
becomes $r_d \sim j^2/(\alpha G M_{\rm BH})$. Thereafter, the accretion
rate on to the BH is given by $\dot{M}_{\rm fb}(t) (R_s/r_d)^s$ and
tracks the mass fall-back rate fairly well.

If the mass fall-back is turned off (say at time $t_1$), the jet power
will revert to the asymptotic scaling $L_{\rm jet}\propto
t^{-4(s+1)/3}$ (eq. \ref{mdotbh1}).  However, there are two distinct
sub-cases possible.  (i) If $t_{\rm acc}\approx 2j^3/(\alpha G^2
M_{\rm BH}^2) \ll t_1$, the jet power will first undergo a sharp drop by a
factor of $t_1/t_{\rm acc}$ on a timescale of $t_1$ (see
eq. \ref{Mdt}) and will only then settle down to the asymptotic
$t^{-4(s+1)/3}$ decline.  (ii) On the other hand, if $t_{\rm acc}
\ \gta\ t_1$, the plateau will smoothly transition to the asymptotic
decline without an intermediate sharp drop.

Fig. 7 shows $L_{\rm jet}$ for three different models of the extended
fall-back, as detailed in the caption.  We see that the jet power
follows closely the time dependence of $\dot{M}_{\rm fb}$ as long as
$\dot{M}_{\rm fb}\not=0$. For instance, when $\dot{M}_{\rm fb}\propto
t^{-0.4}$, $L_{\rm jet}\propto t^{-0.4}$ as well, and so is the case
when the fall-back declines as $t^{-1.2}$. This means that, if the
observed x-ray plateau were to arise due to central engine activity,
we require a long lasting, nearly constant, mass-fall-back rate,
perhaps something like $\dot {M}_{\rm fb}\sim t^{-0.5}$.

\begin{figure}[h!]
\begin{center}
\includegraphics[height=4.5in,width=5.95in]{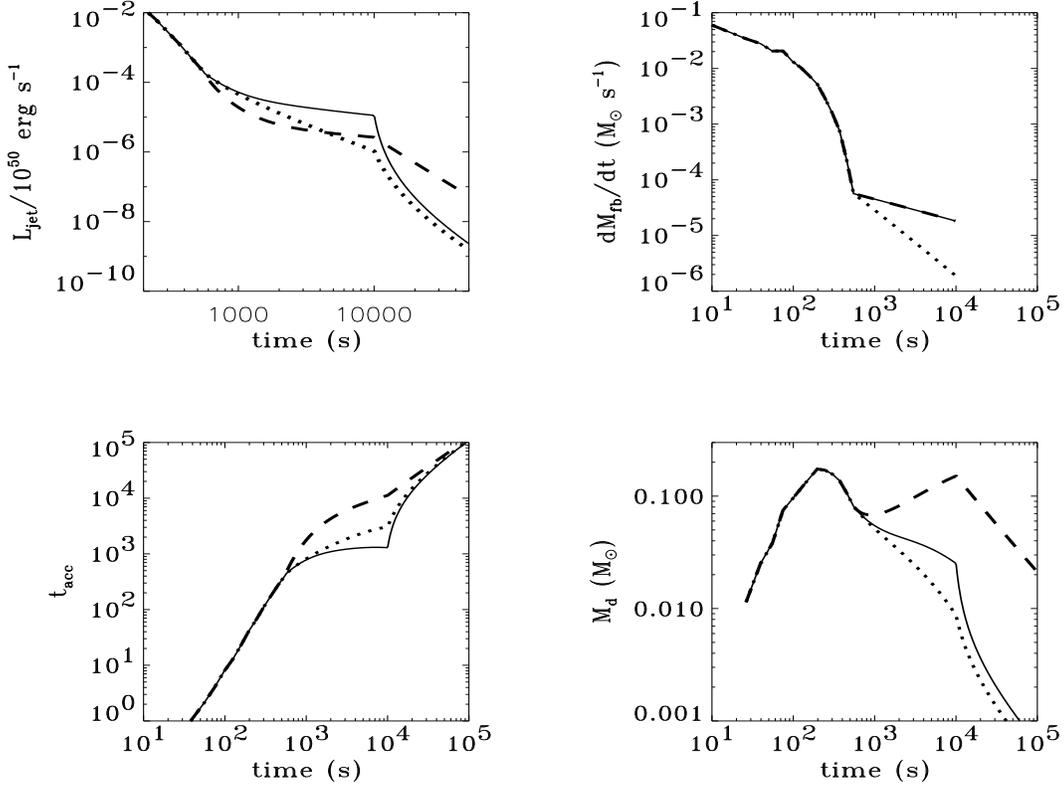}
\caption{\small {\bf Top left:} Shows jet power versus time for three
  different models.  The solid and dashed lines correspond to
  $\dot{M}_{\rm fb}\propto t^{-0.4}$, with $j =2.8\times10^{18}$
  cm$^2$s$^{-1}$ and $8.4\times10^{18}$ cm$^2$s$^{-1}$, respectively.
  The dotted line corresponds to $\dot{M}_{\rm fb}\propto t^{-1.2}$
  with $j =2.8\times10^{18}$ cm$^2$s$^{-1}$; these large values of 
  specific angular momenta correspond to surface layers of the progenitor
   star.  All the models assume
  $\alpha=0.1$, $s=0.75$, $t_1=10^4$ s (end of fall-back), and
  correspond to the stellar model shown in Fig. 2.  Note the sharp
  drop in $L_{\rm jet}$ at the end of the plateau in the solid curve.
  {\bf Top right:} Shows the mass fall-back rate $\dot{M}_{\rm
    fb}$ corresponding to the three models.  {\bf Bottom left:} Shows
  the dependence of the accretion time $t_{\rm acc}$ as a function of
  time.  {\bf Bottom right:} Shows the mass contained in the disk as a
  function of time.}
\label{fig7}
\end{center}
\end{figure}

The second interesting result is that the behavior of the lightcurve
at the end of the plateau depends sensitively on the angular momentum
of the fall-back material.  This is to be expected, of course, since
the angular momentum determines the fall-back radius and thereby the
accretion time.  When the angular momentum is small (solid line in
Fig. 7) we have $t_{\rm acc} \ll t_1$.  This causes a dramatic drop in
the jet luminosity as described above for case (i).  Such a model
can explain GRBs like 060413, 060607A, 070110 (Liang et al. 2007),
which show a sudden drop at the end of the plateau.  On the other
hand, when the angular momentum is larger (dashed line), there is a
smooth transition from the plateau to the asymptotic $t^{-4(s+1)/3}$
tail (case ii), as is seen in other bursts.

\subsubsection{Two scenarios for continued fall-back}

The discussion in the previous subsection was couched in terms of a
parameterised model of the mass fall-back.  Here we consider two
specific scenarios that might produce continued fall-back of gas.

Our first scenario invokes a progenitor star with an outer envelope
which extends out to $R_*\ \gta\ 2\times10^{11}$ cm.  We assume that
this envelope survives the initial implosion/explosion of the star.
The size of the envelope is dictated by the requirement that we must
have continued fall-back until $t \sim10^4$ s.  Thus, the dynamical
time has to be $\sim 10$ times longer than at the outer edge of the
Woosley \& Heger (2006) star, which means that the envelope must have
a radius about a factor of 4 larger than the outer radius of their
model.  Moreover, the star should have a density structure similar to
their model inside $r\sim 5\times10^{10}$ cm, including a very steep
density gradient from $r\sim 3-5\times10^{10}$ cm in order to produce
the rapidly falling lightcurve at the end of the prompt $\gamma$-ray
emission.  The envelope must thus be a distinct entity sitting on top
of the Woosley \& Heger star, and must havce a flatter density profile.

How flat must the density profile in the envelope be?  In \S~2.4, we
derived a scaling for the mass fall-back rate on to the central disk,
which gives $\dot M_{\rm fb}\propto t^0$ ($t^{-1}$) for
$\rho(r)\propto r^{-2}$ ($r^{-3}$); see equation (\ref{mfb3}).  If we
assume for simplicity that the jet power is proportional to $\dot
M_{\rm fb}$, then a shallow plateau light curve $\sim t^{-0.5}$
requires an envelope density profile substantially shallower than
$r^{-3}$.  Actually, the accretion rate on to the black hole is
smaller than $\dot M_{\rm fb}$ by a factor $r_{\rm fb}^{s} \propto
t^{8s/3}\Omega^{2s}$ (eqs. \ref{NDAF}--\ref{ADAF} \& \ref{mfb3}).  If
$r_{\rm fb}$ increases with time, as is likely unless the envelope has
a constant specific angular momentum, then we will need an envelope
with density going as $r^{-2}$ or even shallower.  In any case, the
envelope must rotate rapidly enough to form a centrifugally supported
disk when the gas falls back.

The amount of He in the 16T1 pre-collapse stellar model of Woosley \&
Heger (2006) is 0.37$M_\odot$ and it is concentrated near the surface. A
somewhat more massive and extended He envelope is what is needed
if we want continued fall-back lasting for $\sim10^4$s to produce an
x-ray plateau. 
We note that the wind mass-loss rates from massive stars
are uncertain. Some observations suggest that the typically 
assumed mass-loss rates are too high by a factor of a few 
(Smith et al. 2007; Smith 2007). The implication being that massive stars 
may retain a small fraction of their envelope -- perhaps more often than 
generally assumed. Woosley \& Heger (2006) suggest that their 16T1 model is 
likely to produce a SN Ic upon collapse, and a slightly more extended
envelope probably would not modify that conclusion. 

One of the most attractive features of the envelope scenario is that
it naturally produces a sudden and dramatic drop in the fall-back rate
when the outermost layers of the envelope have fallen back.  Then,
depending on whether $t_{\rm acc}$ is smaller or larger than $t$,
which is determined by the angular momentum of the fall-back gas
(\S~3.5.2), we can have either a large drop in the luminosity (solid
line in Fig. 7) or a smooth roll-over to a power-law tail (dashed
line).  Thus, the model may be able to accommodate most observed
plateau lightcurves.

Our second scenario involves a bona fide supernova explosion in which
a substantial fraction of the outer layers of the star is accelerated
outward on a relatively short time scale.  However, some of the
envelope material fails to escape to infinity and is accreted on to the
BH.  This material contributes to an extended episode of fall-back and
causes a plateau in the light curve.  This model is again subject to a
number of requirements.  A total fall-back mass of $\sim0.5 M_\odot$
is needed in order to explain the x-ray lightcurve during the plateau.
The requirement on the time dependence of $\dot{M}_{\rm fb}$ is fairly
stringent -- it should be no steeper than $t^{-0.5}$, which is a
serious problem.  The canonical fall-back rate of marginally bound
ejecta is $t^{-5/3}$ (Chevalier 1989); a very similar time dependence,
$\dot{M}_{\rm fb}\propto t^{-1.7}$, was also seen in a 1-D collapsar
simulation by MacFadyen et al. (2001).  In addition, it is very
difficult to see how one could have fall-back stopping abruptly, as
required to explain the sudden drop in luminosity at the end of the
plateau in a few bursts (e.g., Troja et al. 2007).  It would require
the layers that fall back to have a density profile varying as $\sim
r^{-2}$ (to reproduce the shallow light curve) out to a radius of
$\sim2\times10^{11}$ cm (to fit the plateau time scale), and then to cut-off
abruptly, presumably because everything outside of this radius escapes
to infinity.

\subsection{Flares}

A major feature of many GRB lightcurves is the presence of one or
more x-ray flares.  Assuming a flare corresponds to a sudden increase
in the jet luminosity, we see from equation (\ref{Lj}) that we require
either $\eta_j$ or $\dot{M}_{\rm BH}$ to change suddenly.  We have
assumed that $\eta_j$ is determined primarily by the BH spin, which is
not likely to change abruptly.  Therefore, we require a sudden burst
in the mass accretion rate.  One possibility is a viscous instability
in the disk (Piran, 1978). The other is a sudden enhancement in the mass
fall-back rate, e.g., if the material in the stellar envelope is held
up for a while by the mass outflow from the disk and then suddenly
finds a way to accrete.

Even if there is an abrupt jump in $\dot{M}_{\rm fb}$, the accretion
rate $\dot{M}_{\rm BH}$ will still be smoothed on the accretion time
scale $t_{\rm acc}$.  However, as Figs. 6 and 7 show, $t_{\rm acc}/t$
is often less than unity.  Therefore, sudden changes in the jet power
are possible in this model.

Yet another possibility is a gravitational instability in the disk.
There could be substantial mass in the reservoir, and this gas could
become self-gravitating and go unstable.  A gravitational instability
could, in principle, produce features in the lightcurve on time scales
faster than the viscous time $t_{\rm acc}$.

A final possibility is that a strong magnetic flux may accumulate
around the black hole during the accretion and may then repeatedly
stop and restart the accretion, causing flares in the x-ray omission
(Proga \& Zhang 2006).  This model is close in spirit to the
``Magnetically Arrested Model'' described by Narayan, Igumenshchev \&
Abramowicz (2003).

\subsection{Hypernova}

A feature of our model is that there is no conventional supernova
explosion.  Except for some material along the polar axes which may
be punched out by the jets, the rest of the stellar mass
falls back either directly into the black hole or on to an accretion
disk.  So how do we explain the supernova-like optical lightcurves
that have been seen in a few GRBs (eg. Hjorth et al. 2003; Stanek et
al. 2003; Modjaz et al. 2006)?

Even though we do not have the usual bounce and outgoing shock that
are present in neutron-star-forming supernovae, our fall-back model does
have mass and energy flowing out of the system.  As we discussed in
\S~3.1, only a fraction of the fall-back mass accretes on to the BH.
The rest is ejected in a disk wind.  Kohri et al. (2005) discussed the
possibility that this wind might boost the energy output of a normal
supernova and perhaps convert a failed supernova to a successful one; 
we note that MacFadyen \& Woosley (1999) had found that for $\alpha\sim 0.1$,
dissipation in the disk can power an energetic disk ``wind'' with enough
$^{56}$Ni loading that it would be supernova-like in its properties.
Here we suggest that, in the case of collapse to a black hole, the
disk wind is the primary source of both mass and energy output from
the system.

For the canonical $14 M_\odot$ stellar model we described earlier, the
black hole ends up with a mass of $\sim10M_\odot$ after all the mass
has fallen back, which means that $\sim4M_\odot$ is ejected in the
disk wind.  To estimate the energy carried away by the wind, we use
the following approximate formula from Kohri et al. (2005) for the
wind luminosity
\begin{equation}
L_w \approx {s\over 2(1-s)}\, {\eta_w\dot{M}_{\rm fb}c^2\over\rfb^s}
\left( {1\over r_{\rm in}^{1-s}}-{1\over\rfb^{1-s}}\right),
\end{equation}
where $r_{\rm in}$ is the inner radius at which the ADAF phase of
accretion ceases, and $\eta_w$ is an efficiency factor for the wind
which probably lies in the range $0.1-0.3$.  Of the three regimes of
accretion described in \S~3.1, regime I with a pure NDAF is not
relevant since there is no mass loss from the disk.  In
the case of regime III, which is a pure ADAF, we set $r_{\rm
in}=R_{\rm isco}$, while in regime II we have
\begin{equation}
II.\qquad \log r_{\rm in} = {1\over(1-s)}\left[\log\mfb + s\log R_s
-s\log\rfb+2.5\right].
\end{equation}
For our canonical model, we estimate the total energy carried away by
the wind to be $2\times10^{52}$erg, which is not dissimilar to the estimated 
explosion energy in hypernovae associated with GRBs (eg. Nakamura et al. 2001,
Mazzali et al.  2003). Computing the optical lightcurve is beyond the scope 
of this paper.  However, we note that the hot gas in the disk wind is likely
to undergo nuclear reactions of various kinds. Also, radioactive decay of 
some of the synthesized elements could produce a supernova-like lightcurve 
in the optical; Woosley has suggested that production of $^{56}$Ni in 
the disk wind could give rise to a bright SNa 
(MacFadyen \& Woosley, 1999; Pruet et al. 2002; see Woosley \& Bloom, 2006, 
for a review).  

\section{Discussion}

This work was motivated by two basic questions posed by the extensive
and excellent Swift observations of GRBs: (1) Why does the
$\gamma$-ray/x-ray flux undergo a sharp decline (flux decreasing as
$t^{-3}$ or faster) about one minute after the start of the burst,
even though the central engine itself is apparently active for hours?
(2) After the sharp decline, how does the power from the engine remain
nearly constant for a period of $\sim10^4$ s to produce the long
plateau that is observed in the x-ray lightcurves of many GRBs?

We have attempted to answer these questions in the context of an
accretion model of GRBs in which the central engine is powered by
accretion on to a BH, and the GRB luminosity is proportional to the
power in the resulting relativistic jets.  We employ the pre-collapse
stellar model of Woosley \& Heger (2006), with the density structure
and rotation profile shown in Fig. 2.  We investigate the
post-collapse accretion activity of this model from $\sim$1 s to
$\sim10^5$ s and compare the model predictions with GRB observations.
We compute the mass in-fall rate during the collapse of the massive
star, and the viscous accretion rate on to the central BH using ideas
from Narayan et al. (2001) and Kohri et al. (2005).  We then estimate
the relativistic jet luminosity from the mass accretion rate and the
BH spin, using a prescription proposed by McKinney (2005).

We find that the jet has a luminosity of about $10^{51} ~{\rm erg\,s^{-1}}$, 
lasting for about 10-20s (Fig. 4), which is of order 
the power and duration of a typical long GRB.  The luminosity
arises from the fall-back and accretion of the outer half of the core
of the progenitor star; the material here has sufficient angular
momentum to go into orbit, whereas the material from the inner half of
the core collapses directly to form the BH.  When the outer layers
of the stellar core -- where the density falls off rapidly with $r$ -- are
accreted via the ADAF process, the jet luminosity drops rapidly. This 
provides a straightforward and natural explanation for the steep decline 
of the early x-ray lightcurve observed by Swift. 

Fluence of gamma-ray bursts can be used to constrain the rotation rate in
the core of their progenitor stars to within a factor of 10 or better; too 
large or too small $\Omega$ in the core results in a small jet luminosity 
(\S3.4), and causes the temporal decay of the lightcurve at the end of 
the prompt emission, for a few hundred seconds, to be either too shallow 
or too steep (fig. 5).

In our model, the plateau in the x-ray lightcurve requires continued
accretion to power the jet.  One possibility is that the viscous
time scale in the disk is so long that it takes a time $\sim10^4$ s
for the material in the disk to accrete.  A viscosity parameter
$\alpha\ \lta\ 10^{-2}$ is required.  In this scenario, although mass
fall-back ceases in a few hundred seconds, accretion continues on for
a few hours (Fig. 6).  While the model succeeds in producing an extended
plateau, it causes the fall-off of the early GRB lightcurve to be less
steep -- more like $t^{-2}$ than $t^{-3}$ -- (Fig. 6) in conflict with
observations.  The decline at the end of the plateau is also fairly
shallow, inconsistent with observations of some GRBs.

A second and, in our opinion, more likely scenario is that there is
continued mass fall-back for the entire duration of the plateau
(Fig. 7).  In this case, the accretion time scale itself is short
(i.e., $\alpha$ is large), so the lightcurve reflects the mass
fall-back rate.  One possibility is that the fall-back is due to
material that fails to be ejected by the supernova explosion.  This
idea has two problems.  First, it is hard to see why there should be
an extended period of a fairly constant fall-back rate as required by
the observations, whereas we expect the fall-back rate to vary much
more steeply as $t^{-5/3}$ (Chevalier 1989).  Second, it is hard to
see how we can have a sudden cutoff in the fall-back at the end of the
plateau as seen in several GRBs.

Another possibility is that the progenitor star has a core-envelope
structure, with the core producing the early GRB and the envelope
producing the plateau.  The core must have a radius of $\sim {\rm
  few}\times10^{10}$ cm to explain the duration of the GRB and the
envelope must have a radius of $\sim {\rm few}\times10^{11}$ cm to
explain the duration of the plateau.  There should be a large density
(or $j$) contrast between the core and the envelope, in order to explain the
sharp cutoff of the prompt emission, and the density profile in the
envelope must be fairly shallow, $\rho\sim r^{-2}$, in order to obtain
a shallow plateau.  Depending on the rotation profile of the star,
various kinds of lightcurves --- including ones in which we have a
very rapid cutoff of the x-ray luminosity at the end of the plateau
--- are possible.  This scenario is therefore capable of explaining
almost all observed cases.  GRBs that do not have a plateau in their
lightcurve are also easily explained; these presumably had progenitors
with only a core and no envelope (like the model shown in Fig. 2).

A nice feature of this model is that we could, in principle, use GRB
observations to deduce the density and rotation structure of the
progenitor star.  On the other hand, it is not clear that evolved
massive stars do have the kind of core-envelope structure we need to
explain a typical GRB x-ray plateau.  (We are not aware of any
pre-supernova models in the literature with the required properties.)

An important implication of the accretion model of GRB central engines
is that the accretion flows are advection-dominated and thus have
strong outflows/winds. We have estimated the total energy in the wind
for the pre-collapse stellar model of Fig. 2 and find it to be about
2x10$^{52}$ erg.  This is sufficient to explode the star, and
might explain the observed energetics of supernovae associated with
GRBs.

A generic prediction of the late fall-back model for the x-ray plateau
is that brighter GRBs should have a weaker (lower luminosity) and
shorter duration plateau.  The reason is that a stronger GRB, with its
stronger jet and wind, is likely to eject more of the stellar envelope
during the main burst. Indeed, recent simulations of relativistic
jet-induced supernovae support this prediction, with more luminous
explosions expelling more of the stellar envelope and leaving less
material available for accretion (eg. Tominaga 2007).  

In order to test this prediction, we have looked at a sample of bursts
with known redshift and isotropic equivalent luminosities (Butler \&
Kocevski 2007).  From this set of bursts, we have selected two
subsets, those with distinct x-ray plateaus (GRBs 070110, 060614,
050315, 060607A, 060729, 070810A) and those clearly lacking x-ray
plateaus (GRBs 071020, 070318, 061007, 050922C, 050826,
070411). Consistent with the prediction of the fall-back model, we
find that the average peak isotropic equivalent luminosity (per
frequency interval) of the subset of bursts with distinct x-ray
plateaus is $\sim$ 4 times lower than that of the subset of bursts
lacking an x-ray plateau.  Moreover, using the same sample of bursts,
we find that the average peak isotropic equivalent luminosity (per
frequency interval) of bursts with an x-ray plateau is lower than that
of bursts without an x-ray plateau at the $>$ 10\% level of
significance, assuming normal distributions for the luminosities of
each of these populations of bursts.

A related prediction is that the plateau should be absent, or at least
weak, in those cases where we see a bright supernova event associated
with a GRB.  The idea is that a bright supernova implies powerful
ejection, and there should be less material available for
accretion.  We have only one well observed case of a GRB-supernova
association in the Swift sample of bursts (GRB 030329), and the x-ray
lightcurve did not have a plateau.  While this observation is
consistent with the late fall-back model, we note that it is just a
single object and therefore the result is not very significant.

It is interesting to note that x-ray plateaus are not seen for short
duration GRBs (see Nakar, 2007, for an excellent review).  This is
consistent with our model.  If short GRBs are the result of the merger
of double neutron star binaries (the currently popular model), then
there is no material in an extended envelope in the progenitor to
produce late fall-back.

The model described in this paper is obviously incomplete.  We have
not provided any quantitative explanation for the x-ray flares seen
during the plateau phase (and even later) in many GRBs.  The only
qualitative idea we have offered is that the flares reflect an
instability in the accretion disk.  We have also simplified the model
considerably by postulating a direct proportionality between the jet
power at the point where it is launched from the BH and the observed
luminosity.  Several factors could seriously modify this relation.
First, the tunneling of the jet through the stellar material may be
inefficient, and so the power that escapes from the surface of the
star may be a small (and variable) fraction of the jet power at the base.
Second, the efficiency with which the escaping jet power is converted
to radiation (the physics of which is poorly understood) may be
variable.  Finally, the beaming of the jet may be different for the
prompt GRB and the x-ray plateau, and may even vary during the
plateau.  We have ignored these complications in the interests of
simplicity.

\noindent{\bf Acknowledgment:} We thank Stan Woosley for providing his
stellar model and for very useful comments on the draft. We are grateful 
to Craig Wheeler for a number of excellent suggestions that improved the 
paper. This work is supported in part by a NSF grant (AST-0406878),
and NASA Swift-GI-program.

\end{document}